\documentclass[10pt]{article}

\usepackage[top=0.85in,left=1.25in,footskip=0.75in]{geometry}

\usepackage{graphicx}
\usepackage{dcolumn}
\usepackage{array}
\usepackage{bm}
\usepackage{amsfonts}
\usepackage{fixltx2e}

\usepackage{amsmath,amssymb}
\usepackage{changepage}
\usepackage[utf8x]{inputenc}
\usepackage{textcomp,marvosym}
\usepackage{cite}
\usepackage{nameref,hyperref}
\usepackage{microtype}
\DisableLigatures[f]{encoding = *, family = * }
\usepackage{array}
\newcolumntype{+}{!{\vrule width 2pt}}
\newlength\savedwidth

\newcommand\thickhline{\noalign{\global\savedwidth\arrayrulewidth\global\arrayrulewidth 2pt}%
\hline
\noalign{\global\arrayrulewidth\savedwidth}}

\usepackage[aboveskip=1pt,labelfont=bf,labelsep=period,justification=raggedright,singlelinecheck=off]{caption}

\makeatletter
\renewcommand{\@biblabel}[1]{\quad#1.}
\makeatother

\date{\today}

\begin{document}
\vspace*{0.2in}

\begin{flushleft}
{\Large
\textbf\newline{Fractal Analyses Reveal Independent Complexity and Predictability of Gait.} 
}
\newline
\\
Dierick Fr\' ed\' eric\textsuperscript{1,2},
Nivard Anne-Laure\textsuperscript{1},
White Olivier\textsuperscript{3,4},
Buisseret Fabien\textsuperscript{1,5,*}
\\
\bigskip
\textbf{1} Forme et Fonctionnement Humain Research Unit,
Department of Physical Therapy, 
Haute Ecole Louvain en Hainaut (HELHa), 
Rue Trieu Kaisin, 136, 
6061 Montignies-sur-Sambre,
Belgium
\\
\textbf{2} Facult\'{e} des Sciences de la Motricit\'{e}, Universit\'{e} catholique de Louvain, Louvain-la-Neuve, Belgium
\\
\textbf{3} Universit\'{e} de Bourgogne INSERM-U1093 Cognition, Action, and Sensorimotor Plasticity,
Campus Universitaire, BP 27877, 21078 Dijon, France.\\
\textbf{4} Acquired Brain Injury Rehabilitation Alliance, School of Health Sciences, University of East Anglia, Norwich, Norfolk, UK.\\
\textbf{5} Service de Physique Nucl\'{e}aire et Subnucl\'{e}aire,
Universit\'{e} de Mons, UMONS  Research
Institute for Complex Systems, Place du Parc 20, 7000 Mons, Belgium \\

\bigskip

* buisseretf@helha.be

\end{flushleft}

\section*{Abstract}

Locomotion is a natural task that has been assessed since decades and used as a proxy to highlight impairments of various origins. Most studies adopted classical linear analyses of spatio-temporal gait parameters. Here, we use more advanced, yet not less practical, non-linear techniques to analyse gait time series of healthy subjects. We aimed at finding more sensitive indexes related to spatio-temporal gait parameters than those previously used, with the hope to better identify abnormal locomotion. We analysed large-scale stride interval time series and mean step width in 34 participants while altering walking direction (forward vs. backward walking) and with or without galvanic vestibular stimulation. The Hurst exponent $\alpha$ and the Minkowski fractal dimension $D$ were computed and interpreted as indexes expressing predictability and complexity of stride interval time series, respectively. We show that $\alpha$ and $D$ accurately capture stride interval changes in function of the experimental condition. Walking forward exhibited maximal complexity ($D$) and hence, adaptability. In contrast, any perturbation (walking backward and/or stimulation of the vestibular system) decreased it. Furthermore, walking backward increased predictability ($\alpha$) through a more stereotyped pattern of the stride interval and galvanic vestibular stimulation reduced predictability. The present study demonstrates the complementary power of the Hurst exponent and the fractal dimension to improve walking classification. These holistic indexes can easily be interpreted in the framework of optimal movement complexity. Our developments may have immediate applications in rehabilitation, diagnosis, and classification procedures.

\section*{Introduction}
The stride interval of normal human walking is the time period between consecutive heel strikes of the same foot \cite{haus96}. For more than two decades, a line of research focused on the understanding of the nature of the subtle variations observed in stride intervals and the origin of typical long-range structures in these variations. Today, these investigations are of paramount importance since they could provide a better understanding of the physiological mechanisms involved in normal human walking and in alterations observed in clinical practice. The nature of these stride interval variations could arise either from noisy neural processes that result in errors in the motor output or from alterations in the motor command that account for balance instabilities \cite{kurz12}. 

Normal gait is characterized by the presence of autocorrelations in the stride interval when considering walking on a sufficiently long time scale \cite{haus95,haus96}. The origin of these autocorrelations may be attributed to neural central pattern generators (CPGs) \cite{haus95,haus96} or a super CPG coupled to a forced Van der Pol oscillator \cite{west03}, and/or to the biomechanics of walking \cite{gate07,ahn13}. For many years, gait analysis has been studied with classical methods adopting biomechanical models in which variability was not of interest. More recent techniques derived from chaos theory are well adapted to analyse time series that exhibit long-range autocorrelation. Importantly, they treat variability as a meaningful interpretable signal. Since the pioneering works of Hausdorff et al. \cite{haus95,haus96}, long-range autocorrelations in time series are estimated by the Hurst or fractal exponent ($\alpha$). A fractal, introduced in 1975 by the French mathematician Beno\^{i}t Mandelbrot (1924--2010) \cite{man75}, is defined as a geometrical structure that has a regular or an uneven shape repeated over all scales of measurement. It is characterized by a fractal dimension ($D$) greater than the spatial dimension of the structure \cite{rand10}. A famous example of such object is a snow flake. Objects that are statistically self-similar -- parts of it show the same statistical properties at many scales -- exhibit strong autocorrelation. The Hurst exponent $\alpha$ is a statistical measure of long-term memory of time series (see \textit{e.g.} \cite{kantz} for a review) and is usually associated to fractal-like behaviour. In particular, the peculiar behavior of the stride interval may be referred to as ``fractal behavior" \cite{haus95}. 

The theoretical model of optimal movement complexity \cite{ster11} is based on the complementary concepts of \textit{predictability} and \textit{complexity}. Nature let us find optimal behavior in terms of skills and variability through evolution. The optimal state of a biological system is characterized by chaotic temporal variations in the steady state output that correspond to maximal predictability. Too few practice results in high disorder (randomness, no predictability) and excessive practice leads to high order (periodic signal, maximal predictability). Adaptation of a system to external stimuli is maximal only at an intermediate state of predictability. Furthermore, a signal from a dynamical system also holds some inherent complexity. A decrease of complexity of a physiological system results from either a reduction in the number of structural components or an alteration in the coupling function between these components. For instance, a joint can become rigid with senescence, hence decreasing the degree of freedom of the system and consequently, its complexity. A holistic approach to study these mechanisms requires to associate specific measurements to these two concepts. The Hurst exponent $\alpha$ captures part of the story and is well suited to reflect predictability. The Minkowski fractal dimension $D$ provides good measurability of the “apparent rugosity” of fractals \cite{falc95} and reflects complexity. Here, we use these parameters to complement the usual quantification of autocorrelation $\alpha$ in unusual and perturbed gait conditions in an attempt to probe adaptability in the framework of the model of optimal movement complexity \cite{ster11}.

 As of today, the vast majority of studies explored autocorrelation in the stride interval during natural forward walking. In one notable exception however, Bollens et al. \cite{bol13} also tested backward walking in a small sample of young healthy subjects. The authors did not find significant differences in long-range autocorrelation between both walking directions. However, backward walking measures revealed to be more sensitive than forward walking measures to classify elderly fallers compare to non-fallers \cite{frit13}. The study of backward walking under the perspective of fractal analyses is therefore promising to provide more reliable predictive index of fallers, as previously proposed for forward walking \cite{haus07}. Backward walking is also frequently used in sports and in rehabilitation settings, and a better understanding of the variability of stride interval in this condition is needed since it is believed that backward walking is at least partly controlled by specialized neural circuits \cite{hoog14}. 

The vestibular system provides an essential sensory contribution to the maintenance of balance during human walking \cite{daki13}. Individuals with vestibular disorders show a decreased walking stability accompanied by an increased risk to fall \cite{whit00}. Therefore, perturbing the vestibular system of healthy subjects with galvanic vestibular stimulation (GVS) is a well targeted mean to probe gait: it is standardized, well tolerated by subjects, generated by currently affordable electrostimulators, and easy to implement when a large number of stride intervals are recorded with an instrumented treadmill. The use of GVS is also an increasingly common clinical intervention on locomotion \cite{buzz03,ding00,ding06}.

Previously, autocorrelations in stride interval time series have been identified not only in healthy young adults \cite{haus95} but also in children \cite{haus99} and elderly \cite{haus97}, and even -- although significantly modified -- in several neurodegenerative conditions. In particular, the cases of Huntington's disease \cite{haus97}, amyotrophic lateral sclerosis \cite{haus00}, and Parkinson's disease have been studied \cite{haus09,war16}, with a hope of connecting the observed modifications of fractal behavior to some relevant evaluation of the risk of falling \cite{haus07}. Here, we hypothesize that the combined effects of walking direction (WD) and the application of GVS on long-range autocorrelations in the stride interval could enhance the sensitivity of fractal analysis to identify impaired gait. We measured $\alpha$  and $D$ during forward and backward walking, with and without the application of binaural and monaural GVS. We speculate that these two indexes should be able to capture differences between experimental conditions and therefore provide better indexes to classify patients.

\section*{Material and methods}
\subsection*{Participants}

Thirty-four undergraduate and graduate healthy students (18 males, 16 females) in physiotherapy took part to this study and were recruited at Haute Ecole Louvain en Hainaut (Charleroi, Belgium). Mean age was 23 years (standard deviation, $SD$=2), height was 173 cm ($SD$=9), mass was 69 kg ($SD$=10), body mass index was 23 kg m$^{-2}$ ($SD$=3), and lower limb length ($L$), measured in standing position as the distance between the floor and great trochanter, was 88 cm ($SD$=5).

Subjects were not medicated and did not exhibit any neuromusculoskeletal, orthopaedic, respiratory, or cardiovascular disorders that could influence their gait. Exclusion criteria included vestibular disorders in addition to specific GVS exclusion criteria: presence of a heart pacemaker, pregnancy, metallic brain implants, epilepsy, and skin damage behind the ears or forehead. Eligible participants were required to be able to respond to verbal questions, comprehend questionnaires, and understand instructions during the procedures of the study. Prior to participating, subjects read and signed an informed consent form. The study was approved by the ethics committee of Grand H\^opital de Charleroi and conducted in accordance with the declaration of Helsinki. 

\subsection*{Experimental procedure}

Subjects walked on an instrumented treadmill (70 cm wide, 185 cm long) with an integrated force plate and an overhead safety frame (N-Mill, Motekforce Link, The Netherlands). They wore comfortable running shoes, a safety harness, and were asked to keep their eyes fixed straight ahead. Four walking conditions were studied during two measurement sessions on two different days: forward walking without GVS (FW$_{S0}$) and with GVS (FW$_{S+}$) and backward walking without GVS (BW$_{S0}$) and with GVS (BW$_{S+}$). Session 1 included FW$_{S0}$ and FW$_{S+}$ conditions and session 2 included BW$_{S0}$ and BW$_{S+}$ conditions. During each condition, subjects walked on the treadmill for 15 minutes. Before each session, subjects were given five minutes to familiarize themselves with the treadmill and the conditions.

\begin{figure}
\caption{{\bf Typical plots of the temporal evolution of the stride interval in the different experimental conditions. FW or BW stand for forward and backward walking respectively. The indices $S+$ or $S0$ indicate the presence or absence of GVS.}
}
\centering
\includegraphics[width=12cm]{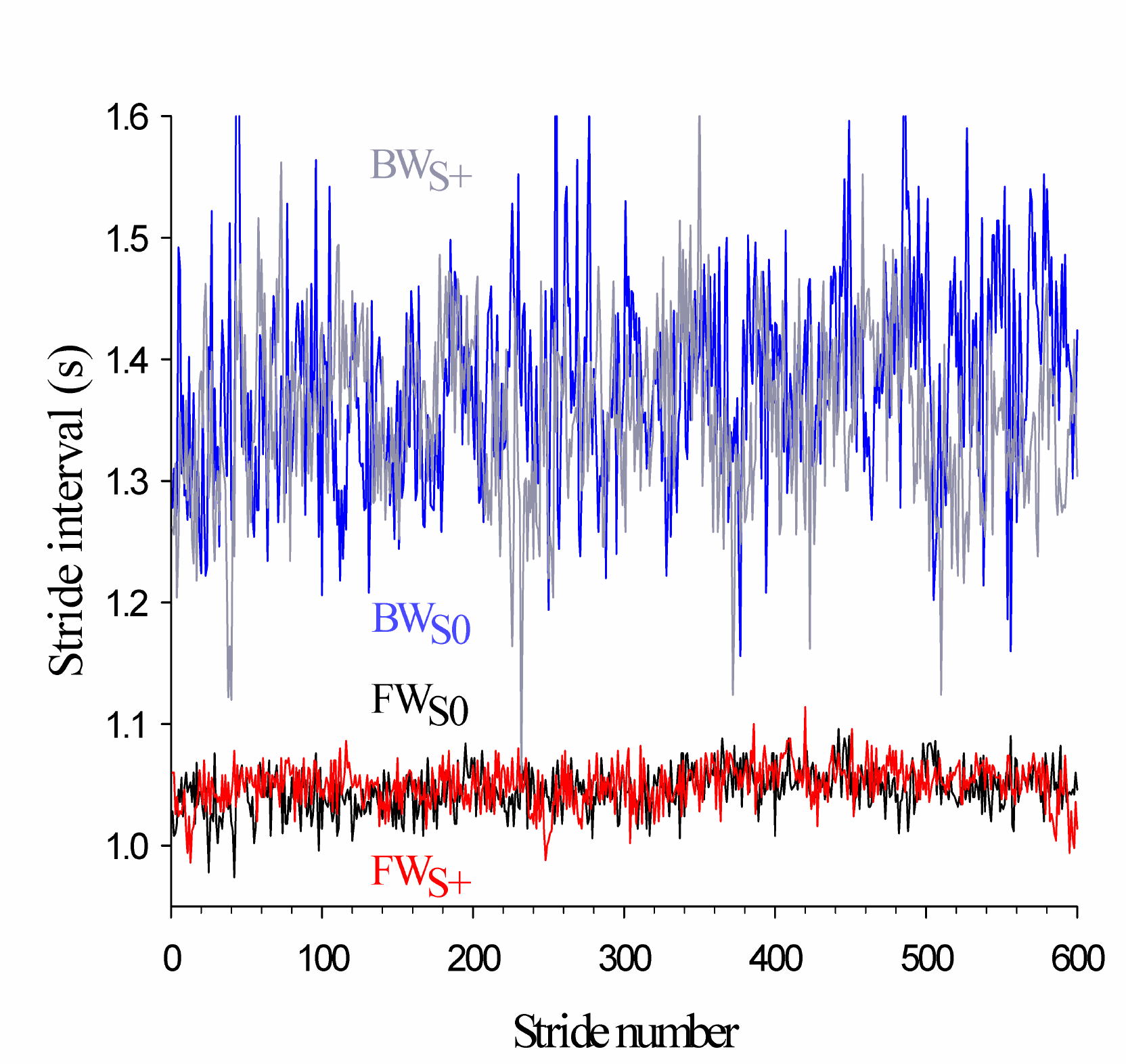}
\label{fig:data}
\end{figure}

Subjects walked at their comfortable speed that was determined during the familiarization procedure by the same experimenter (NAL) by tuning the speed of the belt while the subject was walking without GVS. The same speed was then imposed when GVS was applied. Vertical ground reaction force ($F_v$) and centre of pressure ($CoP$) of each foot was recorded at a sampling rate of 500 Hz using the manufacturer's software (CueFors 2, Motekforce Link, The Netherlands). Time series stride interval were computed from heel strikes (during forward walking) or toe strikes (during backward walking) of the right foot identified on $F_v$-time histories and time series step width from maximal medio-lateral displacement of $CoP$ of two consecutive steps. At the completion of both sessions, four time series containing the values of the stride intervals in the different conditions were obtained for each subject. Typical plots are displayed in Fig. \ref{fig:data}. 

\begin{figure}
\caption{{\bf The three types of galvanic stimulations used.}
}
\centering
\includegraphics[width=9cm]{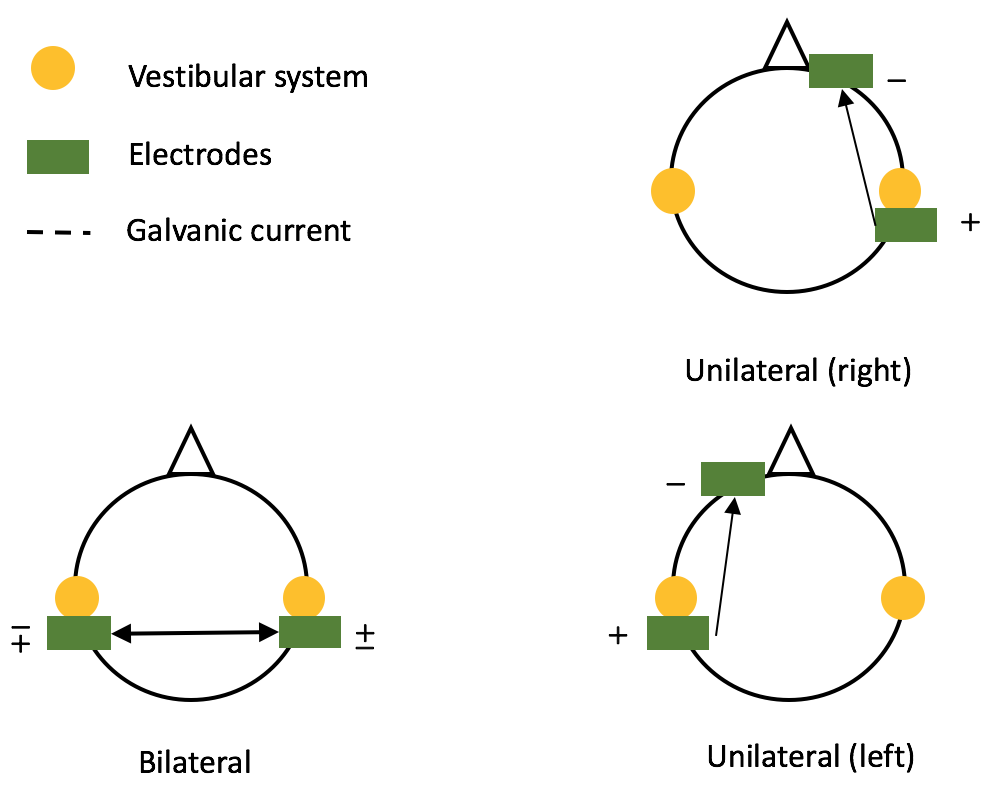}
\label{fig:gvs}
\end{figure}

Bipolar GVS was applied with a regulated, direct-current device (Compex 3 Professional, Compex Medical SA, Switzerland) with a maximum output current of 20 mA by steps of 0.125 mA. The carbon electrodes (20 cm$^{2}$) were covered with a saline-soaked sponge held in place over the mastoids or forehead with a strap. The 34 subjects were randomly exposed to one of the three different transcranial stimulation conditions: binaural ($n=12$), unilateral left ($n=11$), and unilateral right ($n=11$). For the binaural stimulation, electrodes were randomly placed over the mastoids with a cathode-left anode-right or cathode-right anode-left montage. For the monoaural stimulations, the cathode was randomly placed over the right mastoid and the anode on the right part of the forehead (right stimulation) or the cathode was placed over the left mastoid and the anode on the left part of the forehead (left stimulation). Fig. \ref{fig:gvs} shows a schematic representation of the location of the electrodes over the head for the 3 stimulation conditions. The intensity was set at the highest sensory tolerance threshold, that was determined by increasing the current intensity slowly by 0.125-mA steps. That intensity was maintained constant throughout the walking period. Mean current density for subjects was 0.07 mA cm$^{-2}$ (range: 0.04--0.08 mA cm$^{-2}$). The intensity and duration of the GVS adhered to the safety criteria for transcranial direct current stimulation \cite{utz10}. After FW$_{S+}$ and BW$_{S+}$ conditions, each subject completed a home-made French translation of the Motion Sickness Assessment Questionnaire (MSAQ) \cite{gia01}, that consists of 16 questions, allowing to differentiate motion sickness symptoms along the gastrointestinal, central, peripheral, and sopite-related dimensions.

\subsection*{Data analysis}

\begin{figure}
\caption{{\bf Schematic representation of the body (circle) and the lower limb (solid lines) when the heel strikes the ground in forward direction (FW) or the toes in backward direction (BW).}
}
\centering
\includegraphics[width=4cm]{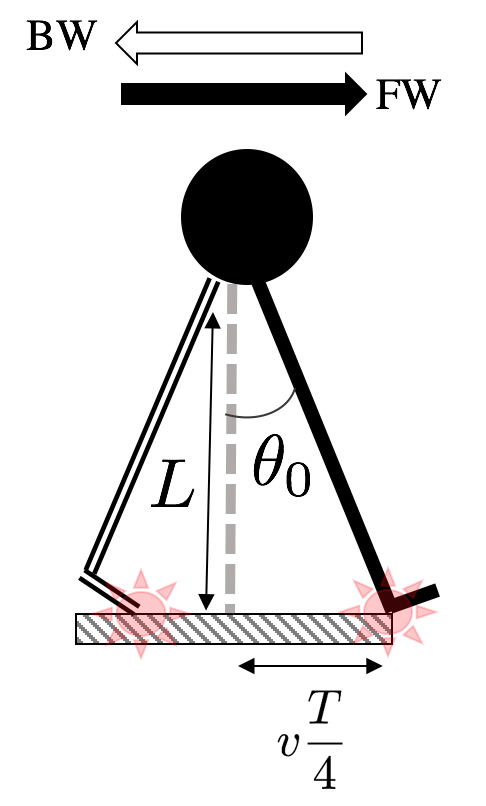}
\label{fig:hs}
\end{figure}

The treadmill software directly computed the mean stride interval, $T$, and the mean step width, $w$, for each subject in each condition. The stride amplitude ($\theta_0$), \textit{i.e.} the angle between the leg and the vertical at heel strike, has then been computed from the relation $v\, \frac{T}{4}=L \tan\theta_0$, displayed in Fig. \ref{fig:hs}, where $v$ is the walking speed and $L$ is the lower limb length of the subject.

The temporal analysis of our experimental data has to go a step beyond mean values to study the information contained in stride interval variability. Let $\bm{ T}=\{ T_i : i=1,\dots,n_0\}$ be a time series, where $T_i$ is the stride interval of cycle $i$ and where $n_0$ is the number of cycles recorded during 15 minutes. The first indicator of variability is the coefficient of variation $CV_T=SD(\bm{ T})/T$. Because $CV_T$ provides no information on the dynamics of the stride interval fluctuations, several indexes characterizing the dynamical structure of the time series have been computed such as the Hurst exponent, $\alpha$, the spectral exponent, $\gamma$, and the Minkowski fractal dimension, $D$. 

We assess the presence of long-range autocorrelations with the Hurst exponent $\alpha$ and the spectral exponent $\gamma$. These parameters quantify the ``predictability" of the time series. As pointed out in \textit{e.g.} \cite{rang00,crev10}, using a single parameter may not be generally enough to assess the presence or not of such autocorrelations. The Hurst exponent, computed by using the Detrended Fluctuation Analysis (DFA) with a linear detrending \cite{peng94}, provides a diagnostic on the long-range trend of the time series. DFA consists in several steps. First, one has to compute the shifted time series $\bm{ T}_\tau=\{ T_{i+\tau} : i=1,\dots,n_0-\tau\}$ and the cumulated time series ${\bm Z}=\{ Z_i=\sum^i_{j=1}(T_j-T) : i=1,\dots, n_0\}$. One has then to divide the cumulated time series ${\bm Z}$ into windows of length $l$, leading to the samples ${\bm Z}^{(m)}(l)$, $m$ labelling the window. For each window, a local least squares linear fit is calculated, leading to the fitted values ${\bm \bar Z}^{(m)}(l)$. Second, one computes the fluctuation function $F(l)=\sqrt{\frac{1}{n_0}\sum^{n_0}_{i=1} \left[Z_i^{(m)}(l)-\bar Z_i^{(m)}(l) \right]^2}$. The Hurst exponent is then defined as the scaling exponent of $F(l)$, \textit{i.e.} $F(l)\propto l^\alpha$. Stationary times series originating from long-range (anti)correlated processes correspond to $0.5<\alpha\lesssim 1$ ($0\leq \alpha < 0.5$), respectively. When $\alpha=0.5$, the process is random. Values larger than 1 correspond to unbounded, unstable, processes \cite{kantz}. In our point of view, a strongly autocorrelated signal can be denoted ``predictable": Its value at a given step is strongly dependent of the system's previous state.

The spectral exponent $\gamma$ can be extracted from the low-frequency behaviour of the power spectral density $P(f)$ of $\bm{ T}$, $f$ being the frequency: $P(f)\propto f^{-\gamma}$. Actually $P(f)$ is the Fourier transform of the autocorrelation function $C(\tau)=E(\bm{ T} \bm{ T}_{\tau})$, where $E$ denotes the average value. The parameter $\gamma$ is expected to take values between 0 and 1 for long-range autocorrelated processes. For large enough time series, the asymptotic relation  
\begin{equation}\label{asrel}
\alpha=\frac{1+\gamma}{2} 
\end{equation}
should relate $\alpha$ and $\gamma$ \cite{kantz}.

We finally compute the Minkowski fractal dimension $D$ of the time series, defined through the box-counting method stating that, if $N(\epsilon)$ is the number of square boxes of size $\epsilon$ needed to fully cover the time series once plotted, then one has $N(\epsilon)\propto \epsilon^{-D}$ for small $\epsilon$. For time series such as the ones we deal with, $D$ will typically lie between 1 (differentiable curve) and 2 (surface with differentiable boundary). Even if $\bm{ T}$ does not define a fractal in a rigorous mathematical way, $D$ can be thought of as a relevant estimator of the ``apparent roughness" of the corresponding curve \cite{gnei11}. In our opinion, assessing the roughness of the stride interval time series -- \textit{i.e.} the variability of fluctuations from one stride to another -- may be associated to the ``complexity" of the process \cite{vail02}. 

It is worth mentioning at this stage that $D$ and $\alpha$ may be seen as independent variables characterizing a time series. The link $\alpha=2-D$ is actually valid only for some widely studied random walks, but processes with arbitrary values of $\alpha$ and $D$ can be built and could be more representative of realistic time series \cite{gnei04}.

\begin{figure}
\caption{{\bf Typical logarithmic plots showing the linear regressions performed to find $\alpha$ (upper left panel), $\gamma$ (upper right panel) and $D$ (lower right panel) in the four conditions. Raw data are those of Fig. \ref{fig:data}. }
}
\centering
\includegraphics[width=14cm]{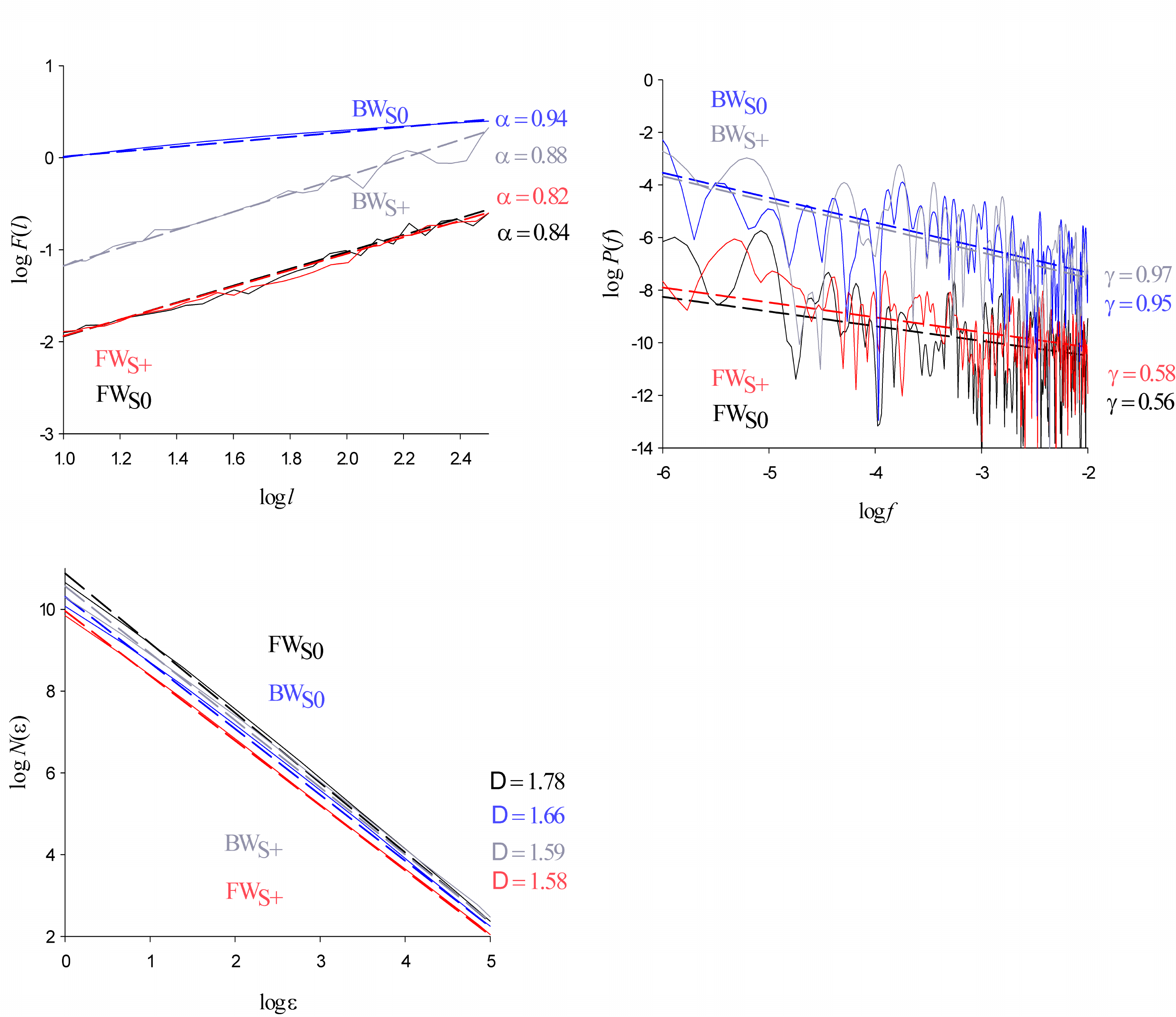}
\label{fig:tech}
\end{figure}

For completeness we show in Fig. \ref{fig:tech} typical plots of the necessary computations performed in view of computing the parameters $\alpha$, $\gamma$ and $D$. All these quantities are actually the slopes of the different linear regressions performed.

\subsection*{Statistical analysis}

All data were checked for normality (Shapiro-Wilk) and equal variance tests. A two-way (WD $\times$ GVS) repeated measures ANOVA with \textit{post hoc} Holm-Sidak method for pairwise multiple comparisons has been performed and used to examine the effects of walking direction ($FW$ or $BW$), GVS ($S+$ or $S0$), and their interaction on the computed parameters. The significance level has been set at $p=.05$ for all analyses and \textit{post hoc} statistical power has been calculated. The correlation between parameters are provided as Spearman's rank correlation coefficient $\rho$, Pearson's correlation coefficient $r$, and prinicipal component analysis (PCA).

Finally, the median ($Mdn$) scores and interquartile ranges ($IQR$) related to the four symptom dimensions assessed by the MSAQ questionnaire have been computed from the 16 items scores according to \cite{gia01}.

All statistical procedures were performed with SigmaPlot software version 11.0 (Systat Software, San Jose, CA). Indexes have been computed using R free software environment (v. 3.2.2) \cite{R}.

\section*{Results} 

\begin{figure}
\caption{{\bf Box plots of the mean stride time ($T$, upper left panel), the coefficient of variation ($CV_T$, upper right panel), the stride amplitude ($\theta_0$, middle left panel), the mean step width ($w$, middle right panel), the Hurst exponent $\alpha$ (lower left panel), and the Minkowski fractal dimension $D$ (lower right panel). }
}
\centering
\includegraphics[width=13cm]{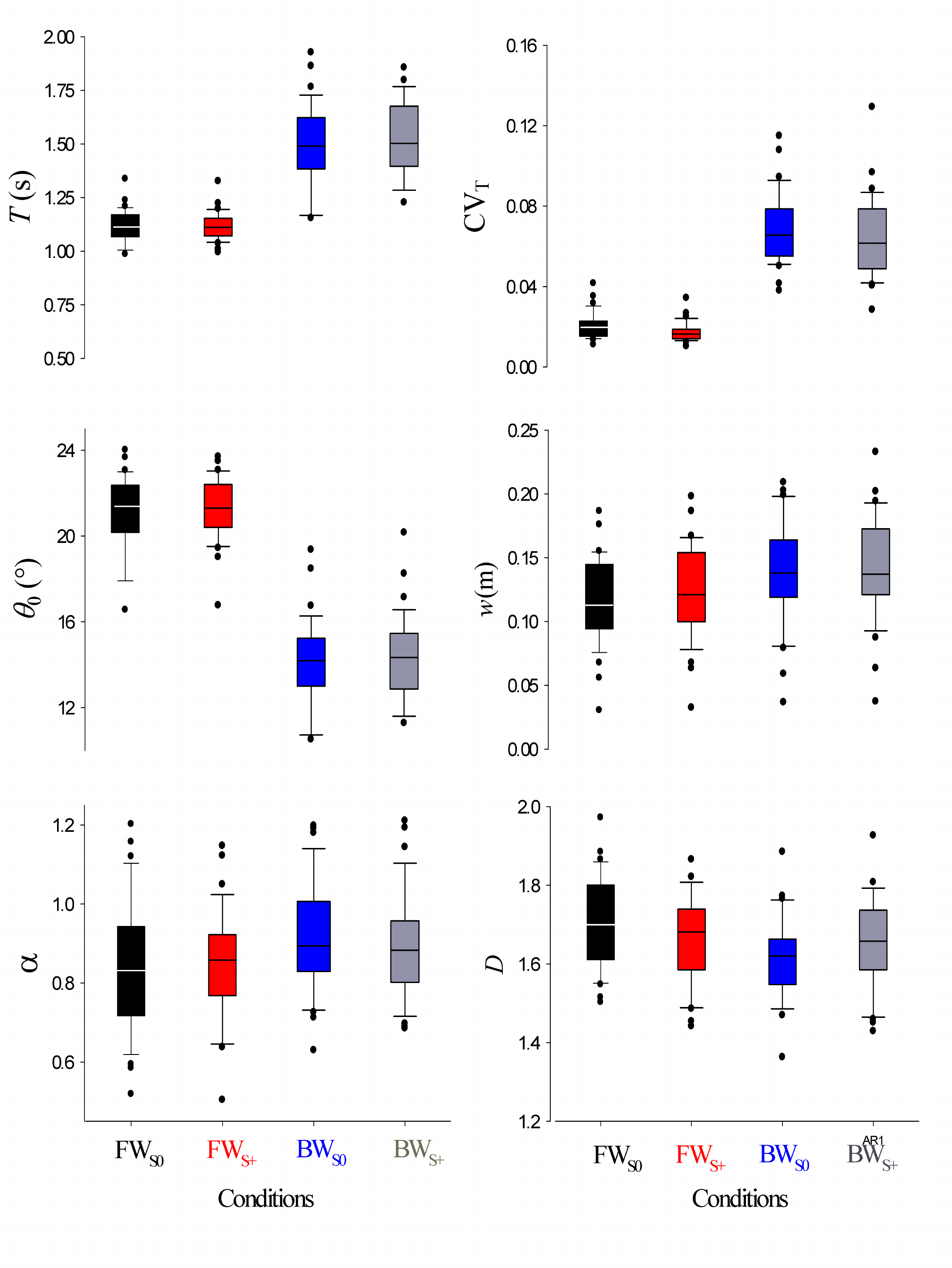}
\label{fig:box}
\end{figure}

Subjects adopted mean comfortable forward speed of 4.4 km~h$^{-1}$ ($SD$=0.4) and backward speed of 2.2 km~h$^{-1}$ ($SD$=0.5). The rather small SDs indicate that our sample was quite homogeneous with respect to that parameter. Results of all parameters analysed according to the four experimental conditions are reported in Table~\ref{tab:res1}. Some parameters particularly relevant for the Discussion are also shown as box plots in Fig. \ref{fig:box}. In any of the four experimental conditions, stride interval time series are such that $0<\gamma<1$ and $0.5<\alpha<1$. These values confirm that time series characterise long-range autocorrelated processes with memory \cite{rang00}.
Note that our value of $\alpha$ in the FW$_{S0}$ condition ($95 \%$ CI of $[0.78-0.90]$) are fully compatible with the interval reported in the meta-analysis \cite{Moo16} in healthy subjects ($95 \%$ CI of $[0.85-0.97]$).
Moreover, Pearson's correlation coefficient between $\alpha$ and $\gamma$ (computed over the complete data set) is equal to $r=0.287$, $p< .01$. Parameters $\alpha$ and $\gamma$ can then be seen as linearly correlated as expected from Eq.~(\ref{asrel}). We decided to keep $\alpha$ since it is the most widely used autocorrelation-related parameter in the literature \cite{Moo16}. The parameter $\gamma$ will not be considered in the following anymore since it only confirms the presence of long-range autocorrelations. In contrast, Pearson's and Spearman's correlation coefficients between $\alpha$ and $D$ are equal to $r=-0.081$, $p=.350$ and $\rho=-0.031$, $p=.719$ respectively. Moreover, a PCA of our complete data set, including the parameters listed in Table  \ref{tab:res1}, shows that nearly 66$\%$ of the total variance is carried by the first two dimensions, the angle between $\alpha$ and $D$ being equal to 131$^{\rm o}$ while that between $alpha$ and $\gamma$ being equal to 14$^{\rm o}$. $\alpha$ and $D$ can then be considered as independent parameters and are both retained in our analysis.

\begin{table}
\begin{center}
\begin{adjustwidth}{-0.5in}{0in} 
\centering
\caption{
{\bf Mean $\pm$ standard deviations values for the parameters computed in the four experimental conditions. FW (BW) stands for forward (backward) walking and S0 (S${\rm +}$) for no GVS (GVS on).}}
\begin{tabular}{l|l|l|l|l}
\hline 
    				 & FW\textsubscript{S0} & FW\textsubscript{S+} & BW\textsubscript{S0} & BW\textsubscript{S+} \\ 
\thickhline		 
    	$T$ (s) & 	$1.12 \pm 0.07$&	$1.11 \pm 0.17 $			&	$1.51 \pm 0.16 $	&		$ 1.55 \pm 0.17 $					\\
$CV_T$ & $0.020\pm0.007$ & $0.017\pm0.006$  & $0.069\pm0.018$ & $0.063\pm0.020$ \\ 
$w$ (s) & $0.104\pm0.029$ & $0.098\pm0.027$  & $0.245\pm0.065$ & $0.244\pm0.065$ \\ 
$\theta_0$ ($^{\rm o}$) & $21.3\pm1.5$ & $21.2\pm 1.4$  & $14.2\pm 1.9$ & $14.5 \pm 1.8$ \\ 
$\alpha$ & $0.725\pm0.062$ & $0.726\pm0.048$  & $0.757\pm0.060$ & $0.716\pm0.064$ \\ 
$\gamma$ & $0.507\pm0.216$ & $0.500\pm0.167$  & $0.655\pm0.194$ & $0.617\pm0.202$ \\ 
$D$ & $1.72\pm0.12$ & $1.66\pm0.11$  & $1.62\pm0.10$ & $1.65\pm0.11$ \\  
\hline
\end{tabular}
\label{tab:res1}
\end{adjustwidth}
\end{center}
\end{table}

\begin{table}
\begin{adjustwidth}{-0.5in}{0in} 

\centering
\caption{
{\bf Significance level ($p-$values) of the \textit{post hoc} (Holm-Sidak) pairwise multiple comparisons between the different experimental conditions for the parameters analyzed. Statistically significant differences are written in bold font.}}
\begin{tabular}{l|l|l|l|l}
\hline  
    				 & FW\textsubscript{S0} vs FW\textsubscript{S+} & BW\textsubscript{S0} vs BW\textsubscript{S+} & FW\textsubscript{S0} vs BW\textsubscript{S0} & FW\textsubscript{S+} vs BW\textsubscript{S+} \\ 
    	\thickhline
 $T$ (s) &      .529          &    	$\bm{ <.001}$	&		$\bm{ <.001}$	& $\bm{ <.001}$\\
$CV_T$ & .308& $\bm{.012}$ & $\bm{ <.001}$ & $\bm{ <.001}$ \\ 
$w$  & $\bm{ <.001}$& .644 & $\bm{ <.001}$ & $\bm{.005}$\\ 
$\theta_0$  & .176 & $\bm{ <.001}$&$\bm{ <.001}$ &$\bm{ <.001}$ \\
$\alpha$ & .857 & .420  &$\bm{ .024}$  & .157 \\ 
$D$ & $\bm{ .042}$ & .143 & $\bm{ <.001}$& .592 \\ 		 
\hline 
\end{tabular}
\label{tab:res2}
\end{adjustwidth}
\end{table}

We first report significant differences found in the two-way RM ANOVA related to WD and GVS factors. WD induces significant differences in all variables (see also Fig. \ref{fig:box}): $T$ ($F$=216, $p< .001$, partial effect size $\eta_p^2$= .991), $CV_T$ ($F$=324, $p< .001$, $\eta_p^2$= .999), $w$ ($F$=13.6, $p< .001$, $\eta_p^2$= .851), $\theta_0$ ($F$=429, $p< .001$ , $\eta_p^2$= .997), $\alpha$ ($F$=5.594, $p= .024$, $\eta_p^2$= .222) and $D$ ($F$=12.1, $p= .001$, $\eta_p^2$= .276). GVS has a less important influence but has nevertheless a significant impact on $T$ ($F$=5.58, $p= .024$, $\eta_p^2$= .135), $CV_T$ ($F$=6.24, $p= .018$, $\eta_p^2$= .872) and $w$ ($F$=11.0, $p= .002$, $\eta_p^2$= .190). The two-way RM ANOVA also reported significant interaction between WD and GVS for $T$ ($F$=1.2, $p= .003$, $\eta_p^2$= .236), $w$ ($F$=4.75, $p= .036$, $\eta_p^2$= .126), $\theta_0$ ($F$=11.4, $p= .002$, $\eta_p^2$= .257) and $D$ ($F$=7.69, $p= .009$, $\eta_p^2$= .189). 

\textit{Post hoc} pairwise multiple comparisons results are listed in Table \ref{tab:res2}. Only $w$ and $D$ are significantly modified when the GVS was active during forward walking. The GVS had a stronger impact in backward walking, with a significant modification of $T$, $CV_T$, $\theta_0$, $\alpha$. The factor having the major impact on the variables is the WD. In FW\textsubscript{S0} vs BW\textsubscript{S0} comparison, all parameters were significantly different. In FW\textsubscript{S+} vs BW\textsubscript{S+} comparison, all parameters were significantly different, except $\alpha$ and $D$. 


MSAQ results are summarized in Table \ref{tab:res3} according to the four dimensions listed in \cite{gia01}. The highest median scores were reached for the items related to the central nervous system dimension. It is worth noticing that beyond these four scores, the third item of the questionnaire \cite{gia01} -- \textit{``I felt annoyed/irritated"} scored highest in the FW$_{S+}$ condition ($Mdn$=4, IQR=$[2-6]$) and in the BW$_{S+}$ condition ($Mdn$=3, IQR=$[1-5]$). 

\begin{table}
\begin{adjustwidth}{-0.5in}{0in} 

\centering
\caption{
{\bf Median scores (in \%) and interquartile range (IQR) results of the MSAQ. The first (Q1) and third quartiles (Q3) are shown between square brackets.}}
\begin{tabular}{l|l|l}
\hline  
    				 & FW\textsubscript{S+} & BW\textsubscript{S+}  \\ 
    	\thickhline
 Central &      27 $[13-51]$         &    22 $[11-40]$	\\
Gastrointestinal & 14 $[11-19]$ & 14 $[11-17]$ \\ 
Peripheral & 11 $[10-15]$ & 11 $[10-19]$ \\ 
Sopite-related & 19 $[14-25]$ & 17 $[11-22]$ \\
\hline 
\end{tabular}
\label{tab:res3}
\end{adjustwidth}
\end{table}

\begin{figure}[ht]
\caption{{\bf Evolution of the Hurst exponent ($\alpha$) versus stride amplitude ($\theta_0$) in the FW$_{S0}$ and BW$_{S0}$ conditions (empty circles). Data from \cite{ahn13} (full circles) are linked with dotted lines for clarity sake. Error bars represent SD. The notation $(NS/^*, NS/^*)$ denotes statistically significant differences ($^*$) or no ($NS$) in the $(\alpha,\theta_0)-$plane. }
}
\includegraphics[width=10cm]{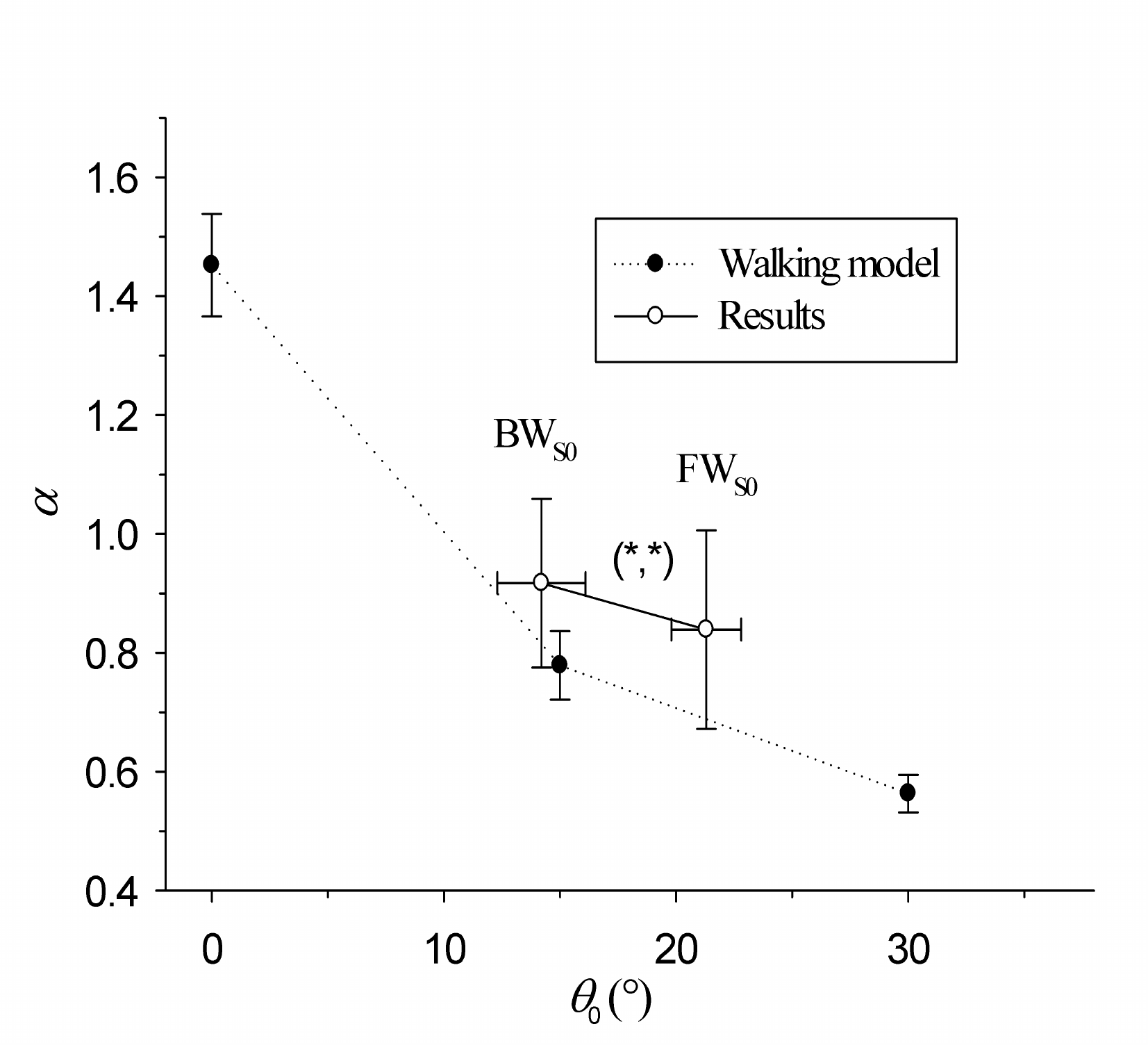}
\label{fig:model}
\end{figure}

\begin{figure}
\caption{{\bf Evolution of the fractal dimension ($D$) versus the Hurst exponent ($\alpha$) taken as complexity and predictability indices respectively for the four experimental conditions. Standard deviations have not been plotted for the sake of clarity. To guide the eyes, arrows indicate the four post-hoc comparisons performed with their significance or not. The notation $(NS/^*, NS/^*)$ denotes statistically significant differences ($^*$) or no ($NS$) in the $(\alpha,D)-$plane. }
}
\includegraphics[width=10cm]{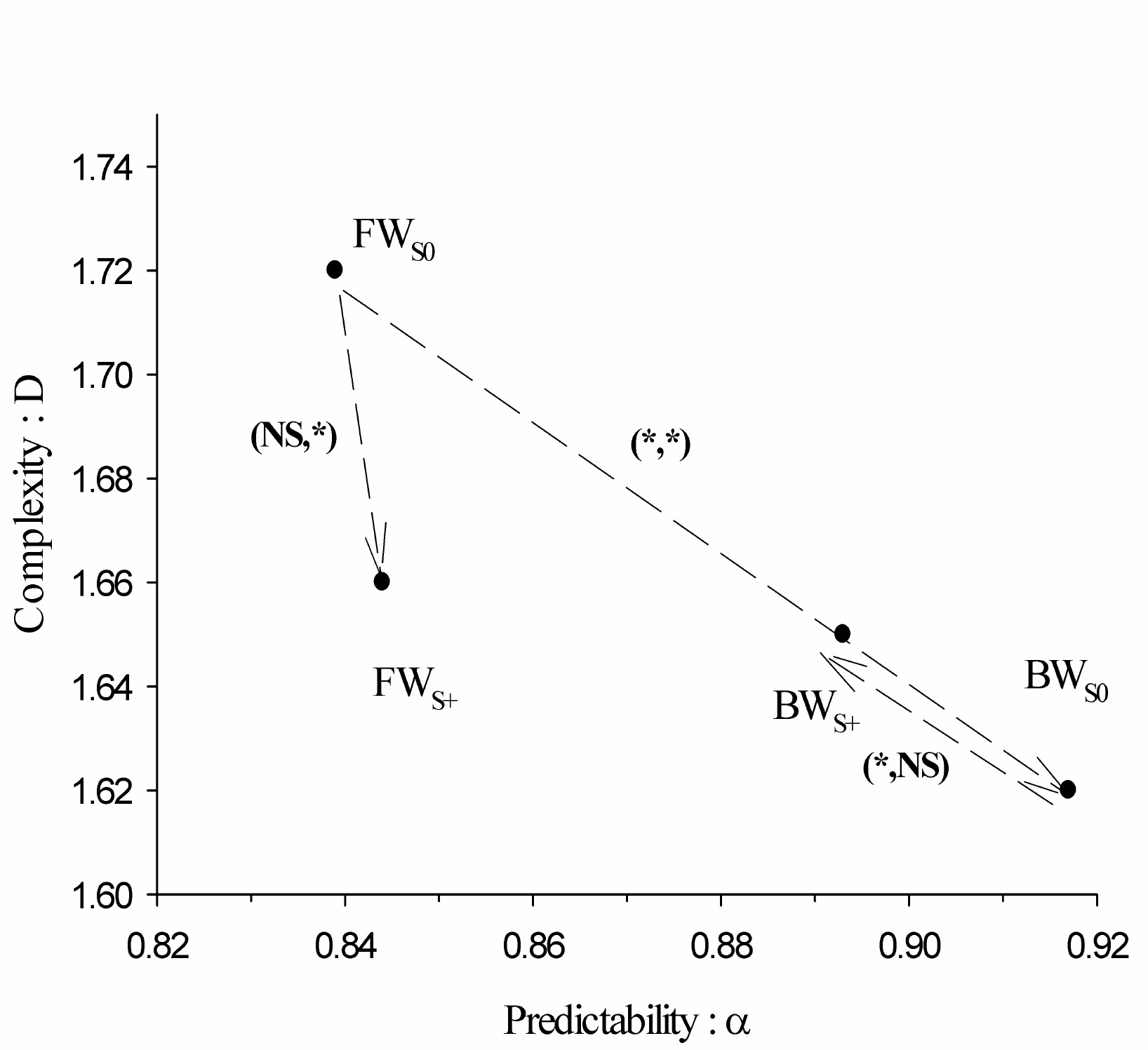}
\label{fig:class}
\end{figure}

Finally, one can ask the question as to how basic kinematic parameters relate to high level indexes such as $\alpha$ and $D$? Spearman's coefficient between $\alpha$ and $\theta_0$, calculated from each subject’s data, is equal to $\rho=-0.207$, $p=.017$. It is therefore relevant to consider that both parameters are correlated. Figure \ref{fig:model} depicts the evolution of $\alpha$ vs $\theta_0$ in the FW$_{S0}$ and BW$_{S0}$ conditions and compares it to that of data presented in Ahn and Hogan’s model (full circles). Figure~\ref{fig:class} reports the evolution of the fractal dimension versus the Hurst exponent in the four experimental conditions. It shows that changing walking direction with respect to the control condition $ FW\textsubscript{S0} $ leads to an increase of $\alpha$ and a decrease of $D$ (independently of GVS). However, turning on GVS during forward walking only leads to a decrease of $D$. There is some decoupling between $\alpha$ and $D$.

\section*{Discussion}

Locomotion has been used as a tool to identify and characterise diverse impairments. Here, we set out to use techniques beyond classical linear analyses of spatio-temporal gait parameters in order to define more sensitive indexes. We asked participants to walk under perturbed conditions induced either by reversing the direction of walking or perturbing the vestibular system, and measured proxies of “complexity” and “predictability” through the Hurst exponent and the fractal dimension.

\subsection*{Walking direction and galvanic vestibular stimulation}

Walking direction has a major impact on our results. Stride interval increased in backward walking compared to forward walking which is in line with previous work \textit{e.g.} \cite{kats10}. This fact is coherent with the lower walking speed spontaneously chosen by the subjects when walking backward. The kinematic parameters $CV_T$ and $w$ are larger in the backward walking conditions than in forward walking. Fluctuations in the stride interval during backward walking are indeed larger than in forward walking. The increase of $CV_T$ has been reported previously \cite{bol13, frit13} in young and elderly adults while, to our knowledge, the corresponding increase in the step width has not been reported elsewhere.

While the existence of autocorrelations in backward and forward walking has been acknowledged in \cite{bol13}, that study did not find a significant difference in the Hurst exponent, presumably because of the smaller sample size (12 subjects). The Hurst exponent appears to be larger in backward compared to forward walking. While neurological diseases generally decrease $\alpha$ \cite{Moo16} (more random motion), an increase in $\alpha$ has been reported in children up to typically 14 years old \cite{haus99}. Both backward walking in adults and forward walking in children can be related to learning processes, with stereotyped, more predictable, motion.

In their recent paper, Ahn and Hogan have shown that long-range autocorrelations may emerge from the dynamics of a particular pendular model of walking described in detail in \cite{ahn12,ahn13}. It is, to our knowledge, the only model linking kinematic variables and autocorrelations indexes. Here, we compare experimental data with that model for the first time. One of the key plots of \cite{ahn13} shows the variation of the Hurst exponent computed on a range of 500 strides for realistic values of the different parameters versus stride amplitude $\theta_0$, which is fixed in their model. A comparison of this plot to our results is shown in Fig. \ref{fig:model}. Results from the FW\textsubscript{S0} condition are in agreement with the model. The BW\textsubscript{S0} condition can be compared too since the dynamical equations presented in \cite{ahn12} are invariant with respect to time reversal. Our results match quite well with the trend of the model. Moreover, GVS has no significant influence on $\alpha$. This is in favour of a mostly mechanical origin of long-range stride interval autocorrelation.

Galvanic vestibular stimulation is a known procedure to electrically stimulate vestibular afferents \cite{scin01,balt04,jahn00}. Here, we used continuous GVS with an average current of 1.4 mA, which was well tolerated by all subjects (low MSAQ median scores).

Previously, vestibular inputs were thought to be primarily required for stabilizing the head to ensure stable gaze control during gait and for spatial orientation in navigational tasks \cite{pozz90,fitz99,jahn00}. More recently, it has also been suggested that vestibular inputs play a role in maintaining dynamic walking stability since they generate phase-dependent influences on lower body control during walking by fine tuning the timing and magnitude of foot displacement \cite{bent04,daki13,blou11}. In agreement with those recent findings, our results show that GVS significantly modifies $T$, $CV_T$ and $w$. The magnitude of $CV_T$ is regularly associated to the risk of fall \cite{vans11}. We observe that turning on GVS significantly decreases $CV_T$ in backward walking; hence training in BW\textsubscript{S+} may be relevant to decrease the risk of fall. Previous research has demonstrated that GVS mostly affects stability in the medio-lateral direction \cite{scin01,balt04,jahn00,kenn05}. This is in accordance with our results that show an increased $w$ during forward walking. 

Interestingly, we found that GVS induced larger $D$ in FW$_{S0}$ compared to FW$_{S+}$ condition. This indicates a less complex stride interval time series in the non-stimulated condition of forward walking. This result is in line with \cite{bent04} showing that planning of the foot placement at heel contact is modulated by vestibular information. Here, we provide another evidence of the influence of GVS on walking variability. It is known that the vestibular system is essential to the maintenance of balance throughout the stepping cycle, with phase-velocity/cadence-dependent modulation on the activity of hip, knee and ankle muscles \cite{daki13}. Vestibular-muscle coupling is specific for each muscle, probably organised according to each muscle's functional role in whole-body stabilization during walking. Our analyses suggest that the less complex nature of the stride interval time series reflects the disruption of dynamic balance evoked by GVS.

\subsection*{The optimal complexity model}

Stergiou and Decker \cite{ster11} proposed that time series originating from human motion could be classified by using two indices. The first catches signal complexity and the other measures its predictability. In this context, a healthy motion should be chaotic, characterized by a maximal complexity reflecting the adaptability of the subject to exterior perturbation, and an intermediate predictability. Pathological motion should be characterized by a lower complexity (fewer adaptability) and a predictability that could be either lower (random, ``drunken-sailor-like", motion) or higher (robotic motion) than the healthy motion. 

Following on that line, we interpret $D$ as a measure of the complexity of gait time series. Indeed, a large fractal dimension is associated to an apparently rough time series, with abrupt relative changes of values stepwise. A complex time series may be the signature of an adaptable behavior. The more a subject is able to change his/her stride interval from one cycle to another, the more s/he should be able to modify his/her pattern. Therefore, $D$ could be a good indicator of complexity during walking. Moreover, we think that the Hurst exponent -- that was independent of $D$ --  could be a relevant predictability index. Indeed, $\alpha$ can discriminate between a random motion ($\alpha=0.5$) and a far more predictable, strongly autocorrelated, time series ($0.5<\alpha\lesssim 1$). So $\alpha$ may provide an answer to the question as to how much a stride interval depends on history? This is exactly what predictability stands for. Healthy subject should be characterised by a maximal value of $D$ (high complexity, good adaptation skills) and an optimal value of $\alpha$ (good but not too high predictability). Any significant deviations from these values could indicate pathological motion linked to any or both dimensions (predictability or complexity).

Our results are displayed in a $(\alpha,D)$-plane in Fig. \ref{fig:class}. It clearly appears that the FW\textsubscript{S0} condition -- the healthy motion -- has the higher complexity and an intermediate predictability as argued in \cite{ster11}. The other conditions, non-standard but not pathological either, have lower complexities. Walking backward without GVS leads to a larger value of $\alpha$, that is a more stereotyped, more predictable walking. GVS slightly decreased $\alpha$ in backward walking. In that condition, walking gets closer to a random process, presumably because of the perturbation of the vestibular system. It is worth noticing that at least one of the two parameters $(\alpha,D)$ is significantly modified when going from one condition to another. A two-dimensional representation is necessary to classify all the experimental conditions we study. Hence it can be conjectured that the only study of $\alpha$ in pathological cases may be too restrictive to discriminate between the pathologies and that more non-linear indexes are worth to be added. 

Previous studies only computed the Hurst exponent and implicitly considered that the fractal dimension and the exponent were related. Here, we computed $D$ beside $\alpha$ and show for the first time that these two parameters are actually decoupled in some conditions. As already pointed out, children walking forward also have a larger $\alpha$ than healthy young adults walking forward. We have calculated from a freely available dataset \cite{haus99} the average $D$ for the 50 children having participated to the study. We found values equal to $1.45\pm 0.21$, a significantly lower value than in our FW$_{S0}$ condition ($t=-6.77$, $p<.001$), as expected. In Parkinson’s disease, $\alpha$ are smaller than in young healthy adults. It has been shown \cite{war16} that $\alpha$ decreases with disease's severity. We have computed $D=1.69\pm0.10$ from the data of \cite{war16} (20 patients with Parkinson's disease walking for 10 min, with $\alpha=0.70\pm0.09$). Although lower than our maximal, FW$_{S0}$, the difference between both values is not significant ($t=-0.941$, $p=.351$). Similarly, neurodegenerative pathologies have actually been shown to generally decrease $\alpha$ with respect to its optimal value \cite{Moo16}.  As can be deduced from the above discussion, the parameters $D$ and $\alpha$ are good candidates to disentangle and characterise the main long-term features of walking. The Hurst exponent $\alpha$ is a widely used indicator of long-term autocorrelations, and adding $D$ opens new classification perspectives. Our findings may have immediate applications in rehabilitation, diagnosis, and classification procedures.

We also think that this field could benefit in a near future from new techniques such as a representation of the stride interval time series in terms of complex networks (visibility graphs) \cite{gao15,gao16}. This technique has already proven to be efficient to distinguish healthy from epileptic EEG signals \cite{gao17}, hence it can reasonably be assumed that visibility graphs could provide relevant information on the structure of stride interval time series. New classification schemes allowed nowadays by machine learning could also shed new light on walking dynamics. Algorithms like random forests could help to find better indices to disentangle the different experimental conditions \cite{ho98}. There is hope that such new techniques could better classify the stride interval time series, but with less common indices, either less intuitive or less easily compared to the literature in walking analysis. Such a research program is beyond the scope of the present study and we leave it for future investigations. 

\section*{Conclusion}

Our findings show that stride interval dynamics behave as a chaotic system exhibiting long-range autocorrelations independently of walking direction. The Hurst exponent is increased when walking backward, suggesting that the more predictable fluctuations of the stride interval reflect more stereotyped motion adopted by subjects in response to this nonstandard condition. The magnitude of these fluctuations are however larger in backward walking, due to the weaker stability of the subjects. The Minkowski fractal dimension complements the characterisation of stride interval variability by considering complexity, or, more intuitively the adaptive capacities of the subject in motion. Any nonstandard condition reduced complexity. The present study thus opens new avenues as to how more accurately classify healthy or pathological walking according to the complexity and predictability of stride interval time series.

\section*{Acknowledgments}

The authors thank C Detrembleur and T Warlop for stimulating discussions at early stages of this work and for having provided us the data of \cite{war16}, and M Scohier for useful comments.

\end{document}